\documentclass[slac_one]{revtex4}
\usepackage{graphicx}
\usepackage{fancyhdr}
\def\units#1{\hbox{$\,{\rm #1}$}}
\def\t#1#2{\hbox{$\theta_{#1#2}$}}
\begin{document}

%Title of paper
\title{The Double Chooz Experiment} %% Paper title goes here

% Repeat the \author .. \affiliation  etc. as needed
%
% \affiliation command applies to all authors since the last
% \affiliation command. The \affiliation command should follow the
% other information

\author{C. E. Lane (for the Double Chooz collaboration)}
\affiliation{Drexel University, Philadelphia, 19104 USA}

\begin{abstract}
The Double Chooz experiment returns to the site of the
Chooz experiment with a pair of detectors for a differential
neutrino flux measurement, providing sensitivity to 
$\sin^2 2\t13 > 0.03$.  Reaching this goal requires
significant improvements in systematic uncertainties, 
based
on the experience with previous reactor neutrino experiments.
\end{abstract}

%\maketitle must follow title, authors, abstract
\maketitle

\thispagestyle{fancy}

% body of paper here - Use proper section commands
% References should be done using the \cite, \ref, and \label commands
% Put \label in argument of \section for cross-referencing
%\section{\label{}}

\section{MOTIVATION}

%Now that neutrino oscillation has been firmly established for
%atmospheric neutrinos\cite{Ashie:2005ik}, solar neutrinos\cite{Ahmad:2002jz}
%and reactor neutrinos\cite{Araki:2004mb}, the next generation of neutrino
%experiments will be targeting the areas in which neutrino
%properties are still unknown: whether neutrinos are Majorana
%or Dirac particles, their mass hierarchy, whether there is
%CP violation in the neutrino sector, and the value of the
%\t13
% mixing angle.
%
Knowledge of the \t13
neutrino mixing angle
is of great importance for many of the goals of
future neutrino research, as described elsewhere\cite{Freedman:2004rt}.  A measurement
of \t13
using reactor anti-neutrinos would be particularly
`clean', with no contamination from
other neutrino flavors, no matter effects, and 
is generally unaffected by other unknown neutrino properties
such as the mass hierarchy.

The current limit experimental limit is $\sin^22\t13 < 0.17$,
depending somewhat on $\delta m^2$, which comes from the
Chooz experiment\cite{Apollonio:2002gd}.  The Chooz experiment used
a single detector, approximately 1\units{km} from a pair
of power reactor cores.  
The Chooz experiment measured a ratio of measured
to expected neutrino events\cite{Apollonio:2002gd}
of $R=1.01\pm 2.8\%(\hbox{\rm stat})\pm2.7\%(\hbox{\rm syst})$.
While the Chooz statistical error is slightly larger, the
sensitivity limit
of the Chooz experiment came from systematic errors; the
experiment was terminated when statistical errors 
reached the same level as systematic errors, so that
little would be gained by continued operation.

\section{CHOOZ AND DOUBLE CHOOZ}

The Double Chooz experiment has the goal of measuring
\t13  for values of $\sin^22\t13 > 0.03$,  
or establishing a limit. It is hoped that such a 
timely measurement will be of use in
the design and prioritization of the next steps in
experimental neutrino physics. 
%
%As implied by the name of the experiment, 
Double Chooz
will return to the Chooz site in northern France, where
two 4.3\units{GWth}
reactors will provide the  anti-neutrino flux. 
The underground laboratory used by the Chooz experiment
($\approx 300\units{m.w.e}$) has been renovated for 
 the new experiment, saving 
cost and time. 
The Chooz experiment had a total systematic 
uncertainty of 2.7\%,  of which 1.6\%
was attributed to absolute detector uncertainty;
%which is
the lowest of any similar
experiment to date.

%To reach the goal 
For Double Chooz,
these uncertainties must be reduced by approximately a 
factor of five.  Most of this improvement will be 
achieved by performing a relative measurement 
between two detectors placed at `near' ($\approx 400\units{m}$)
and `far' (the existing lab, at 1050\units{m}) sites. 
One major goal in the construction of the detectors is to
make them  as identical as possible, however a
% previous reactor neutrino experiment that
Bugey\cite{Declais:1994su}
experiment
attempted to have identical detectors, and found
2\%
relative error, so it is clear that building
identical detectors to within a fraction of a percent is a challenge.  

The Double Chooz detectors will be roughly two times larger 
in target mass than the
original Chooz detector, and will be operated longer with
full reactor power to improve
statistics, yet it is the systematic uncertainty
that will be the limitation.
%
%Even with a relative measurement, there is a significant challenge
%in creating a reactor neutrino experiment with sufficiently
%small systematic uncertainties, but 
It is particularly important 
to have an accurate appreciation of the
sources and size of potential systematics, based
on measurements and experience rather than just simulations. 

\section{EXPERIMENTAL DESIGN}

The Double Chooz detectors\cite{Ardellier:2006mn}
were designed to improve on the Chooz
detector, examining in detail each item in the error budget and 
incorporating ways to reduce or eliminate uncertainty.
Like many previous experiments, Double Chooz will use 
inverse beta-decay from reactor antineutrinos on 
target protons, yielding a `prompt' positron with
energy derived from the original neutrino energy, and an
epithermal neutron which thermalizes and captures
on Gadolinium, giving a `delayed' signal
of approximately 8\units{MeV}.
%
%The neutron capture provides not only a valuable 
%`tag' of neutrino events, with its energy deposition
%well above radioactive backgrounds, but also indicates
%that the neutron capture occurred within the target
%volume of Gd-loaded scintillator.  While there are
%edge effects from neutrons that
%cross the target volume  boundary (which
%requires careful correction for an absolute
%measurement with a single detector) the neutron capture
%on a Gd-loaded target defines a target mass
%with much less uncertainty than techniques using
%reconstructed event positions. 

%\subsection{Detector Siting}

The Double Chooz experiment is being built at the Chooz
nuclear reactor power station, in the Ardennes region of
northern France, operated by \'Electricit\'e de France (EDF).
%Figure~\ref{fig:ChoozMap} shows a map of the reactor site,
%with the location of the two laboratories planned for the experiment.

%\begin{figure}[h]
%\includegraphics[width=4in]{DC_prop_0606025v4.pdf-pg11.eps}
%\caption[Map of Chooz power station]{Map of the Chooz power station. 
%The two reactor cores (indicated by green dots near the 
%center of the figure) are located 140\units{m} apart, and the far
%detector site (marked {\bf FD}) is located 1.0 and 1.1\units{km} from
%the two reactor cores. The currently proposed location for the near
%detector site (marked {\bf ND}) is also shown. 
%}
%\label{fig:ChoozMap}
%\end{figure}

\begin{table}[h]
\begin{tabular}{|c|c|l|c|}
\hline
Systematics&Type&{Chooz}&2 Identical Detectors\cr
           &    &      &Low Background\cr
\hline
&Neutrino Flux and Cross-section&1.9\%&{$\cal O$}(0.1\%)\cr
Reactor&Thermal power&0.7\%&{$\cal O$}(0.1\%)\cr
&E/fission&0.6\%&{$\cal O$}(0.1\%)\cr
\hline
\multicolumn{2}{|c|}{SUM}&2.1\%&{$\cal O$}(0.1\%)\cr
\hline
\end{tabular}
\caption{Comparison of Chooz(absolute) and a two-detector (relative)
 reactor-based error budgets}
\label{tab:ReactorErrors}
\end{table}

A significant amount of the systematic uncertainty in
the Chooz result came from `reactor-related' factors,
such as the absolute reactor thermal power output (see Tab.~\ref{tab:ReactorErrors}). 
Using two detectors to make the neutrino measurement 
relative helps to remove these errors arising from reactor
sources.  
%For example, while there is systematic uncertainty
%in the neutrino cross-section, the cross-section affects
%both detectors of a relative measurement approximately the same.
%Similarly for the reactor thermal power, burn-up, fission yield, etc.
%
For a single reactor core, these reactor-related
effects cancel, 
%since the detectors will receive
%neutrinos from the same source, 
even
as the reactor power level and fuel composition changes.
The Chooz power station has two reactor cores, each
of approximately 4.3\units{GWth} at full power.   
Ideally, the
two detectors should be sited so that the distances to the
two reactor cores are in the same ratio for both the near
and the far detectors.  The distance ratio is not 
exactly identical for Double Chooz ($465\units{m}/351\units{m}$ for the
near detector {\it vs.} $1114.6\units{m}/997.9\units{m}$ for the
far detector), because of practical
issues in the siting of the near laboratory, but  
there still is substantial cancellation of reactor-related 
systematics to acceptable levels.
The ability to cancel reactor-related
uncertainties by matching distance ratios only works for one or two
reactor cores.  More reactor cores gives more overall neutrino
flux, but also more systematic error as the reactor cores contribute
different (relative) amounts of neutrino flux to the detectors. 
An additional advantage to using only two reactor cores is that
it results in a better ability to measure non-reactor backgrounds
during the reactor fueling cycle, by comparing
rates during full-power operation to rates during shutdowns.  
%Reactor operators attempt
%to only shut down one core at a time, resulting in a 50\%
%change in neutrino flux for two cores, but significantly less
%for a larger number of cores.  
To the extent that background
(cosmic-ray or radioactivity) contribute to the systematic
uncertainty, the ability to get a clean measurement of the
background is very desirable. 

%\subsection{Detector Volumes}

\begin{figure}
\begin{center}
\includegraphics[height=2.5in]{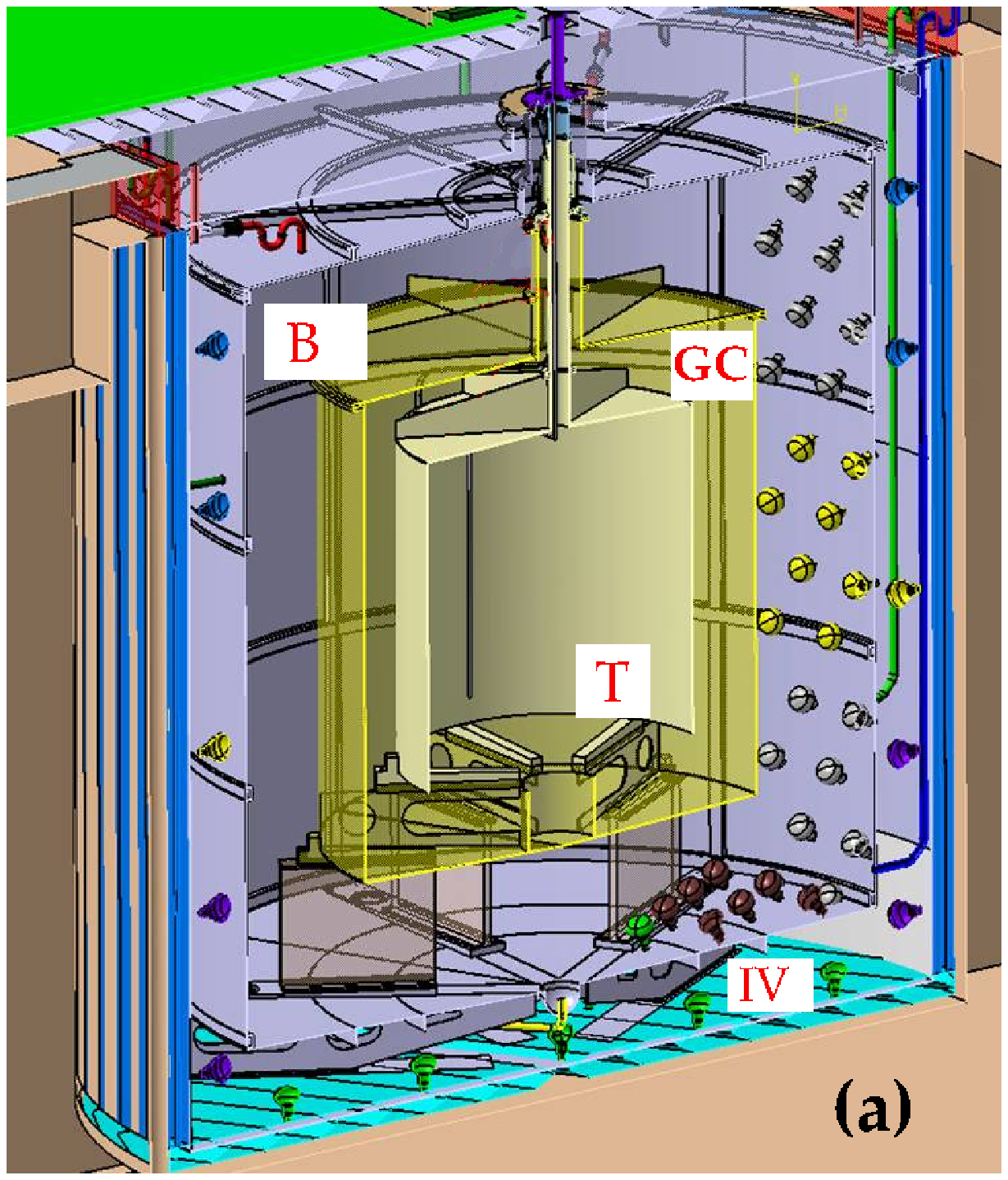}
\includegraphics[height=2.5in]{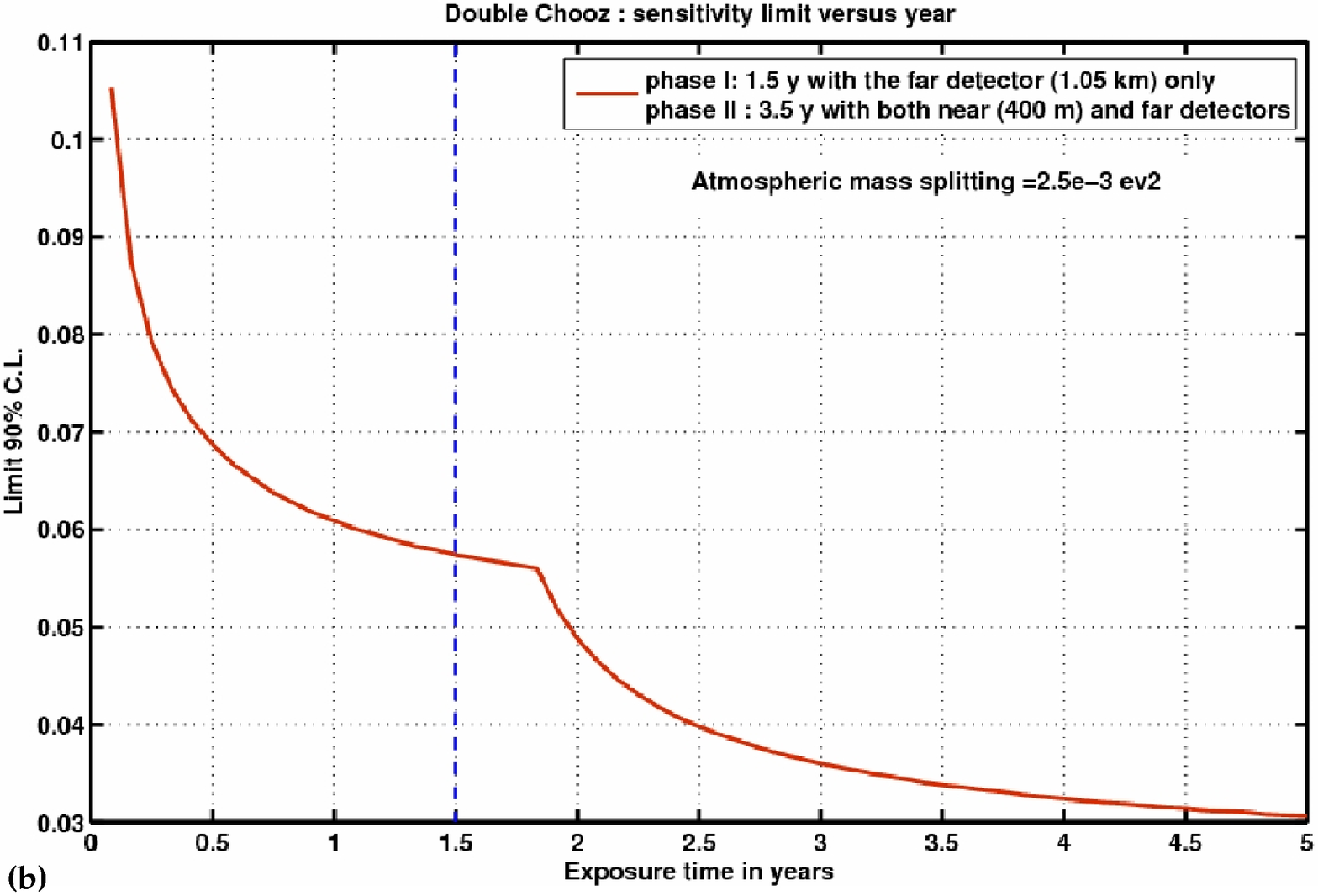}
\caption[Schematic view of detector]{{\bf(a)} Schematic view of the Double Chooz detector design. The target volume
 {\bf T} contains Gd-loaded scintillator,  
 surrounded by a gamma-catcher ({\bf GC}) with  non-loaded
scintillator for collecting the energy of Gd capture events in the target. Outside
the {\bf GC} is a non-scintillating buffer liquid ({\bf B})  
to shield
the inner volumes and suppress background events near the PMTs.  
The PMTs (390 Hamamatsu R7081's) that collect light from the inner detector volumes are located inside
the buffer {\bf B}, and mounted on a stainless steel tank which provides some shielding
and optical isolation.
Outside the stainless steel tank is an Inner 
Veto {\bf IV}, with non-loaded
scintillator and PMTs.
% The purpose of the IV is to efficiently tag cosmic-rays
%penetrating the detector, while further shielding the inner detector from
%gammas and neutrons from outside the detector.  The IV is contained within
%steel shielding that attenuates radioactivity from outside the detector.

{\bf (b)} Double Chooz $\sin^22\t13$ limit as a function of 
time after starting operation.  For the first (approx.) 1.8 years only
the far detector will be running, while the near detector is built, but
even as an `absolute' measurement there will be considerably improvement
in sensitivity to \t13
over previous experiments. After the start of the near detector, the
sensitivity will improve further as a relative measurement can be performed,
reducing the systematic uncertainties.
}
\label{fig:sens_v_t}
\label{fig:detector}
\end{center}
\end{figure}

Figure~\ref{fig:detector}(a)
shows a schematic view of a Double Chooz detector.  Like the Chooz detector,
there is an inner volume of Gd-loaded scintillator which defines the 
neutrino target, surrounded by a volume of non-Gd-loaded 
scintillator for improved capture energy resolution.
%
%(the `gamma catcher') to detect gammas escaping from the target
%volume and improve energy resolution.
%, particularly gammas from neutron capture on Gd.  
%This
%improves the energy resolution of the neutron capture signal, and
%allows energy cuts on the neutron capture signal to be placed where
%there they have minimal systematic effect. 
%
%This improves the energy resolution of the neutron capture signal, and
%allows energy cuts to be placed in a region well separated from the 
%capture signal, thus minimizing systematic effects.
%
%
%Both the target and the gamma-catcher are contained in 
%acrylic vessels, so scintillation light is efficiently
%transmitted to  PMTs.  
%The inner-detector is in
%a stainless steel tank, with a non-scintillating buffer between the 
%tank and the gamma-catcher. There are
%390 specially developed low-background inner-detector PMTs 
%(10-inch, Hamamatsu model R7081) mounted 
%on the stainless steel tank, pointing inward,
%immersed in the buffer liquid.
%
%A  buffer shields the active regions of the detector 
%against gammas and neutrons
%from outside, but
%also suppresses events close to the PMTs; such events
%have poor energy resolution, and tend to contaminate the neutrino
%sample. 
Surrounding the stainless steel tank is an 
``Inner Veto'' detector, consisting of
non-loaded liquid scintillator with 78 Hamamatsu 8-inch R1408 PMTs that
were originally used for the IMB experiment\cite{BeckerSzendy:1992hr},
then for the Super-K\cite{Fukuda:2002uc} veto, 
and have been refurbished for Double
Chooz. The purpose of the Inner Veto is to tag cosmic-ray muons
very efficiently, so that they can be vetoed in analysis.  
The Inner Veto is surrounded by steel shielding to attenuate
radioactivity entering the detector. 
There is also a cosmic-ray muon tracker system (not shown in the 
detector figure) 
that will be installed above
the detectors, so that muon-related backgrounds can be 
more precisely measured.

%\subsection{Detector Uncertainties}

One simple example of a detector uncertainty is from the
mass of protons in the target volume.  For the Chooz experiment,
this particular uncertainty was a 1\% 
effect on the overall error.
%and since Chooz was an absolute measurement, the uncertainty
%contributed to the overall experimental error.  
%Part of this 1\%
%was caused by uncertainty in measurement of the target volume, but
%part was caused by uncertainty in the target scintillator H/C ratio (the
%Gd compounds used for loading the scintillator were somewhat hygroscopic,
%leading to uncertainties in their hydration state, and hence uncertainties
%in the amount of hydrogen added to the scintillator).  
%
Double Chooz will do a relative measurement between two 
detectors so
such uncertainties will largely cancel out to the extent that the
detectors can be made identical in construction and operation.  
%Making
%two detectors identical is not trivial; as evidenced by the
%previously cited example\cite{Declais:1994su}
%with a 2\%
%relative error between detectors.
%
The goal for Double Chooz is to achieve a relative uncertainty of 0.6\%
in `detector related' uncertainties between the near and far detectors.
To again take the example of the target  mass: one would like the
target volumes to be constructed with identical volumes, but realistically
the mechanical tolerances, acrylic properties, and other effects indicate 
that the relative uncertainty
may be $\sim 0.6\%$.  To reduce this uncertainty, 
some of the steps Double Chooz will be implementing are: using
several techniques for measuring the target scintillator mass, 
using a scintillator
with well-defined chemical composition (to minimize H/C ratio uncertainty) 
and implementing active thermal measurement and control to ensure that any
differences between the two detector targets are known with relative 
uncertainty $<0.2\%$. 
%
%No significant difference is expected 
%%between scintillators used in the far detector (starting earlier)
%and the near detector.  While it is not yet known if such steps are
%needed, there is the possibility that near- and far-detector scintillators
%can be mixed, or that the `old' far-detector scintillator discarded, so
%that when operating with two detectors the scintillators are 
%very close to identical. 

%Since the Double Chooz detectors are contained within steel 
%outer shielding, 
%there is also
%a possibility that the magnetic field environment could differ between
%the two detectors.  
%To minimize this effect, the shielding steel is 
%demagnetized before construction, mu-metal
%magnetic shields are being used as part of the PMT enclosures, 
%and monitors are being installed in the
%detector to measure the fields and quantify any changes. 

Significant effort is being put toward identifying possible
differences between the two detectors,  finding means to reduce
the differences, and to make measurements that will allow precise 
compensation.  This may seem excessive by the
standards of previous reactor neutrino experiments, but when one 
tries to reduce systematic uncertainties to a fraction of a percent,
a level that has not been previously reached by experiments of this
type, all sources of possible uncertainty must be examined for their
potential effect.

\section{ANALYSIS TECHNIQUES}

The Chooz experiment had 1.5\% 
systematic uncertainty arising from
analysis effects,
required a series of seven analysis
cuts (positron energy, matching positron and neutron positions, etc.)
in order to reduce backgrounds from radioactivity, for
an overall neutrino event efficiency of $\approx 70\%$.
The Double Chooz detector has been designed to
eliminate the problems with radioactivity seen in Chooz,
by adding a buffer volume, and selecting detector components for
low activity. 
% This should result in much less need for analysis
%cuts to remove noise events.  
%
%Each analysis cut adds systematic error, so for
The Double Chooz detector is designed such that only a minimum
number of analysis cuts are needed (see Tab.~\ref{tab:analysis_errors}).
%Improved calibrations will  
%allow
%the use of neutron capture energy cuts that are well separated from
%the 8\units{MeV} capture peak, in a region where the cut efficiency
%is insensitive to the energy threshold. 
%
The result is fewer analysis cuts:
three (or two, with low background)
for Double Chooz data, 
minimizing systematics for each cut,
with
an overall contribution to systematic uncertainty of $\sim0.3\%$.

\begin{table}[h]
\begin{tabular}{lrrc}
\hline
&Chooz&\multicolumn{2}{c}{Double Chooz}\cr
cut&rel. err(\%)&rel. err(\%)&comment\cr
\hline
Positron Energy${}^*$&0.8&0&not used\cr
Positron-Geode distance&0.1&0&not used\cr
Neutron capture on Gd&1.0&0.2&Cf calibration\cr
Capture energy containment&0.4&0.2&Energy calibration\cr
Neutron-Geode distance&0.1&0&not used\cr
Neutron delay&0.4&0.1&--\cr
Positron-Neutron distance&0.3&0--0.2&0 if not used\cr
Neutron multiplicity${}^*$&0.5&0&not used\cr
\hline
Combined&1.5&0.2--0.3&\cr
\hline
${}^*$ average values
\end{tabular}
\caption[Comparison of analysis uncertainties]
{Comparison 
of analysis systematic uncertainties for
Chooz and Double Chooz.
An improved detector design reduces backgrounds,
 allowing the elimination of many
cuts.
%
%The reduction in radioactive background
%in Double Chooz should make it possible to eliminate the cuts
%for `Positron Energy' and `Neutron Multiplicity', and also 
%the position cuts needed to remove
%background events near the edge of the Chooz detector (`Positron-Geode
%distance' and `Neutron-Geode distance') and possibly the position
%correlation between positron and neutron.  Improved calibration and
%separation of the active regions of the detector from the PMTs allow
%reduced systematic uncertainties for `Neutron Capture', `Capture
%energy containment', and `Neutron delay'.
} 
\label{tab:analysis_errors}
\end{table}

To see the development between Chooz and Double Chooz,
a ``Positron Energy'' cut of 1.3\units{MeV} 
was needed in Chooz to reduce
low-energy background, but introduced systematics 
from the knowledge of the threshold, from 
inhomogeneous energy response in the
detector, and from scintillator variation with time, all of which 
contributed to the `Positron Energy' uncertainty
listed above.  
%
%%%
In Double Chooz the 
non-scintillating
buffer and other background reduction measures will greatly 
decrease the rate of low-energy backgrounds, allowing Double Chooz
to use just a hardware energy threshold set well below 1\units{MeV},
and eliminate the ``Positron Energy'' cut and its 
systematic uncertainty.

\section{PROGRESS, TIMESCALE, CONCLUSIONS}

At the time of this writing (Sep. 2008) the Double Chooz far detector
construction has been underway for a few months.  The installation of
the Inner Veto PMTs is expected near the beginning of 2009, followed
by the insertion of the inner stainless steel tank, inner PMTs, acrylic vessels, etc.,
with the `first neutrino event' expected by Summer 2009.
Figure~\ref{fig:sens_v_t}(b)
shows the anticipated sensitivity of Double Chooz to \t13
as a function of time from when the far detector starts operation.

%\begin{figure}[h]
%\begin{center}
%\includegraphics[width=4in]{sensitivity_oxford_0907.eps}
%\caption{
%Double Chooz $\sin^22\t13$ limit as a function of 
%time after starting operation.  For the first (approx.) 1.8 years only
%the far detector will be running, while the near detector is built, but
%even as an `absolute' measurement there will be considerably improvement
%in sensitivity to \t13
%over previous experiments. After the start of the near detector, the
%sensitivity will improve further as a relative measurement can be performed,
%reducing the systematic uncertainties.
%}
%\label{fig:sens_v_t}
%\end{center}
%\end{figure}

While later experiments will no doubt improve the statistics of 
a \t13
search, pushing further downward in \t13
is primarily a matter of reducing the systematics. 
The largest improvement in systematics comes from moving from
an absolute (Chooz) to a relative measurement (Double Chooz).
Operating additional detectors at similar distances will 
allow comparisons that can improve
how systematics are quantified, but it is much less clear
that a simple `scale up' would significantly
improve the measurement of \t13 
unless
the underlying sources of systematic uncertainties are addressed.

\begin{acknowledgments} 
The Double Chooz Collaboration gratefully acknowledges
support from the U.S.A., France, Spain, Germany,
the U.K., Japan and Brazil. 
\end{acknowledgments}

%\subsection{References}
\bibliography{refs}
\bibliographystyle{h-physrev4}

\end{document}